\documentclass[10pt, twocolumn, conference]{IEEEtran}

\usepackage{graphicx}
\usepackage{amsmath,amssymb}
\usepackage{subfigure}
\usepackage{xspace}
\usepackage{psfrag}
\usepackage{bbm}
\usepackage{mathrsfs}

\usepackage{xspace}
\usepackage{bbm}
\catcode`@=11
\chardef\mathlig@atcode\count255

\def\actively#1#2{\begingroup\uccode`\~=`#2\relax\uppercase{\endgroup#1~}}
\def\mathlig@gobble{\afterassignment\mathlig@next@cmd\let\mathlig@next= }
\def\mathlig@delim{\mathlig@delim}
\def\mathlig@defcs#1{\expandafter\def\csname#1\endcsname}
\def\mathlig@let@cs#1#2{\expandafter\let\expandafter#1\csname#2\endcsname}
\def\mathlig@appendcs#1#2{\expandafter\edef\csname#1\endcsname{\csname#1\endcsname#2}}

\def\mathlig#1#2{\mathlig@checklig#1\mathlig@end\mathlig@defcs{mathlig@back@#1}{#2}\ignorespaces}

\def\mathlig@checklig#1#2\mathlig@end{%
 \expandafter\ifx\csname mathlig@forw@#1\endcsname\relax
 \expandafter\mathchardef\csname mathlig@back@#1\endcsname=\mathcode`#1%
 \mathcode`#1"8000\actively\def#1{\csname mathlig@look@#1\endcsname}%
 \mathlig@dolig#1\mathlig@delim
\fi
\mathlig@checksuffix#1#2\mathlig@end
}

\def\mathlig@checksuffix#1#2\mathlig@end{%
\ifx\mathlig@delim#2\mathlig@delim\relax\else\mathlig@checksuffix@{#1}#2\mathlig@end\fi
}
\def\mathlig@checksuffix@#1#2#3\mathlig@end{%
\expandafter\ifx\csname mathlig@forw@#1#2\endcsname\relax\mathlig@dosuffix{#1}{#2}\fi
\mathlig@checksuffix{#1#2}#3\mathlig@end
}

\def\mathlig@dosuffix#1#2{%
\mathlig@appendcs{mathlig@toks@#1}{#2}%
\mathlig@dolig{#1}{#2}\mathlig@delim
}

\def\mathlig@dolig#1#2\mathlig@delim{%
 \mathlig@defcs{mathlig@look@#1#2}{%
 \mathlig@let@cs\mathlig@next{mathlig@forw@#1#2}\futurelet\mathlig@next@tok\mathlig@next}%
 \mathlig@defcs{mathlig@forw@#1#2}{%
  \mathlig@let@cs\mathlig@next{mathlig@back@#1#2}%
  \mathlig@let@cs\checker{mathlig@chck@#1#2}%
  \mathlig@let@cs\mathligtoks{mathlig@toks@#1#2}%
  \expandafter\ifx\expandafter\mathlig@delim\mathligtoks\mathlig@delim\relax\else
  \expandafter\checker\mathligtoks\mathlig@delim\fi
  \mathlig@next
 }%
 \mathlig@defcs{mathlig@toks@#1#2}{}%
 \mathlig@defcs{mathlig@chck@#1#2}##1##2\mathlig@delim{%
  \ifx\mathlig@next@tok##1%
   \mathlig@let@cs\mathlig@next@cmd{mathlig@look@#1#2##1}\let\mathlig@next\mathlig@gobble
  \fi 
  \ifx\mathlig@delim##2\mathlig@delim\relax\else
   \csname mathlig@chck@#1#2\endcsname##2\mathlig@delim
  \fi
 }%
 \ifx\mathlig@delim#2\mathlig@delim\else
  \mathlig@defcs{mathlig@back@#1#2}{\csname mathlig@back@#1\endcsname #2}%
 \fi
}%

\catcode`@\mathlig@atcode

\newcommand{\muspace}{\mspace{1mu}}

\DeclareRobustCommand{\scond}{\mathchoice{\muspace\vert\muspace}{\vert}{\vert}{\vert}}
\mathlig{|}{\scond}

\DeclareRobustCommand{\discint}{\mathchoice{\mspace{-1.5mu}:\mspace{-1.5mu}}{\mspace{-1.5mu}:\mspace{-1.5mu}}{:}{:}}
\mathlig{::}{\discint}

\newcommand{\Ac}{\mathcal{A}}
\newcommand{\Bc}{\mathcal{B}}

\newcommand{\Sc}{\mathcal{S}}

\newcommand{\Xc}{\mathcal{X}}
\newcommand{\Yc}{\mathcal{Y}}

\newcommand{\pen}{{P_e^{(n)}}}

\newcommand{\aep}{{\mathcal{T}_{\epsilon}^{(n)}}}
\newcommand{\aepvar}{{\mathcal{T}_{\epsilon'}^{(n)}}}

\newcommand{\Mh}{{\hat{M}}}

\newcommand{\Sh}{{\hat{S}}}

\newcommand{\kh}{{\hat{k}}}
\newcommand{\lh}{{\hat{l}}}
\newcommand{\mh}{{\hat{m}}}
\newcommand{\sh}{{\hat{s}}}

\newcommand{\zh}{{\hat{z}}}

\newcommand{\Rt}{{\tilde{R}}}

\def\e{\epsilon}

\DeclareMathOperator\E{\textsf{E}}
\let\P\relax
\DeclareMathOperator\P{\textsf{P}}

\DeclareMathOperator\C{\textsf{C}}

\def\error{\mathrm{e}}

\newcommand{\N}{\mathrm{N}}

\def\textiid{i.i.d.\@\xspace}
\newcommand\iid{\ifmmode\text{ i.i.d. } \else \textiid \fi}

\def\mathllap{\mathpalette\mathllapinternal}
\def\mathllapinternal#1#2{%
  \llap{$\mathsurround=0pt#1{#2}$}}

\def\clap#1{\hbox to 0pt{\hss#1\hss}}
\def\mathclap{\mathpalette\mathclapinternal}
\def\mathclapinternal#1#2{%
  \clap{$\mathsurround=0pt#1{#2}$}}

\let\oldstackrel\stackrel
\renewcommand{\stackrel}[2]{\oldstackrel{\mathclap{#1}}{#2}}

\renewcommand{\hbar}{h\mathllap{\overline{\vphantom{h}\hphantom{\rule{4.6pt}{0pt}}}\mspace{0.77mu}}}

\catcode`~=11 
\newcommand{\urltilde}{\kern -.06em\lower -.06em\hbox{~}\kern .02em}
\catcode`~=13 

\begin{document}

\newtheorem{theorem}{Theorem}
\newtheorem{lemma}{Lemma}
\newtheorem{mydef}{Definition}
\newtheorem{corollary}{Corollary}
\newtheorem{example}{Example}
\newtheorem{remark}{Remark}
\newtheorem{proposition}{Proposition}

\newcommand{\CSC}{C_\mathrm{SC}}
\newcommand{\CC}{C_\mathrm{C}}
\newcommand{\CNC}{C_\mathrm{NC}}

\title{Action Dependent Strictly Causal State Communication}

\author{Chiranjib Choudhuri and Urbashi Mitra}
\maketitle
\long\def\symbolfootnote[#1]#2{\begingroup%
\def\thefootnote{\fnsymbol{footnote}}\footnote[#1]{#2}\endgroup}
\symbolfootnote[0]{Chiranjib Choudhuri (cchoudhu@usc.edu) and Urbashi Mitra (ubli@usc.edu) are with the Ming Hsieh Department of Electrical Engineering, University of Southern California, University Park, Los Angeles, CA 90089, USA.}
\long\def\symbolfootnote[#1]#2{\begingroup%
\def\thefootnote{\fnsymbol{footnote}}\footnote[#1]{#2}\endgroup}
\symbolfootnote[0]{This research has been funded in part by the following grants and organizations: 
ONR N00014-09-1-0700,  NSF CNS-0832186, NSF CNS-0821750 (MRI), NSF CCF-0917343, NSF CCF-1117896 and DOT CA-26-7084-00.}


\begin{abstract}
The problem of communication and state estimation is considered in the context of channels with action-dependent states. Given the message to be communicated, the transmitter chooses an action
sequence that affects the formation of the channel states, and then creates the channel input sequence
based on the state sequence. The decoder estimates the channel to some distortion as well as decodes
the message. The capacity--distortion tradeoff of such a channel is characterized for the case when
the state information is available strictly causally at the channel encoder. The problem setting extends
the action dependent framework of \cite{Weissman2010} and as a special case recovers the results of
few previously considered joint communication and estimation scenarios in \cite{Zhang--Vedantam--Mitra2011, Choudhuri--Kim--Mitra2010, Choudhuri--Kim--Mitra2011}. The scenario when the action
is also allowed to depend on the past observed states ({\em adaptive action}) is also considered. It is
shown that such adaptive action yields an improved capacity--distortion function.
\end{abstract}


\section{Introduction}

\par Consider the example scenario of an autonomous underwater vehicle engaged in a classification task
communicating with a surface station; in particular the vehicle employs {\bf active} classification
wherein it controls the views it has of the target (or state, $S$). The vehicle can modify its position,
sensor parameters, etc. One can envision that the vehicle would modify its plan as it collects new
information about the target state \cite{Hollinger--Mitra--Sukhatme2011}. This scenario motivates our
examination of both adaptive and non-adaptive active communication over channels with state. In
particular, we are interested in scenarios where the encoder can select actions (dependent on the
message to be sent) that are potentially dependent on the past channel states in order to communicate
the state as well as additional information to the destination.

\par In this framework, encoding is in two parts: given the message, an action sequence is created. The
actions affect the formation of the channel states, which are accessible to the transmitter in a strictly
causal manner when producing the channel input sequence. A channel with action-dependent states
then is characterized by two ingredients: the distribution of state given an action $p(s|a)$ and, the
distribution of the channel output given the input and state $p(y|x,s)$. We are interested in the
scenario when in addition to communicating pure information across the channel, the transmitter also
wishes to help reveal the channel state to the receiver. We characterize the tradeoff between the
independent information rate and the accuracy of estimation of the channel state via the
\textit{capacity-distortion function} (first introduced in~\cite{Zhang--Vedantam--Mitra2011}). The wide
applicability of our framework can be seen in the following problems which can be expressed as a
problem of conveying action dependent state to the destination: active classification \cite{Naghshvar--Javidi2010}, underwater path planning \cite{Vasilescu--Kotay--Rus--Dunbabin--Corke2005, Hollinger--Mitra--Sukhatme2011}, data storage over memory with defects \cite{Kusnetsov--Tsybakov1974,Heegard--El-Gamal1983}, dynamic spectrum access systems \cite{Haykin2005}, {\em
etc.}.

\par Pure communication over channels with action-dependent states was introduced
in~\cite{Weissman2010} wherein the capacity of such a channel with both non-causal and causal
state information at the encoder were characterized. Due to our goal of acquiring the channel at the
destination as well as information transmission, a distinctly different approach is taken herein relative
to the Gelfand-Pinsker methodology adopted in \cite{Weissman2010}. However, we are able to recover
the results of \cite{Weissman2010} for the causal case (extending the same proof strategy as in strictly causal case), revealing an alternative proof strategy. We observe that the codes which are optimal for achieving capacity may not be good codes for state estimation.

\par Alternatively, our work extends that of joint communication and state estimation in \cite{Sutivong--Chiang--Cover--Kim2005, Zhang--Vedantam--Mitra2011, Choudhuri--Kim--Mitra2010, Choudhuri--Kim--Mitra2011}; conditioned on the action sequence, we have such a problem. The role of the action sequence is to not only communicate the message, but to also determine a good communication channel for both the message as well as the state estimation. The contributions of our work is as follows: we characterize the capacity distortion function for this problem via a two stage encoding scheme. In stage one, information is encoded in the action sequence; in stage two, conditioned on the
action sequence, a block Markov strategy akin to that in \cite{Choudhuri--Kim--Mitra2010} is shown
to be capacity--distortion optimal. We show that strictly causal CSI improves the channel estimate,
while the capacity is unchanged. Our results are generalized to the case where the action sequence is a
function of both the message and the past channel states (feedback), we denote this as {\em adaptive
action}; the benefits of such an encoding are quantified. In addition to the generalizations previously
mentioned ({\em i.e.} \cite{Zhang--Vedantam--Mitra2011, Choudhuri--Kim--Mitra2010, Choudhuri--Kim--Mitra2011, Weissman2010}), we show that our adaptive action framework recovers prior results on
multiple access channels with states~\cite{Lapidoth--Steinberg2010, Li--Simeone--Yener2011}.

\par The rest of this paper is organized as follows. Section II describes the basic channel model with
discrete alphabets, characterizes the capacity--distortion function, establishes its achievability and
proves the converse part of the theorem. Section III extends the results to the adaptive action setting,
wherein we allow the feedback from the past states to the action encoder. Section IV illustrates our
results with few examples. Finally, Section V concludes the paper.


\section{Problem Setup and Main Result}\label{sec:asymmetric-strictly-causal}

We assume a discrete memoryless channel (DMC) with discrete memoryless state (DMS) model $(\Xc \times \Sc \times \Ac, p(y|x,s)p(s|a), \Yc)$ that consists of a finite input alphabet $\Xc$, a finite output alphabet $\Yc$, a finite state alphabet $\Sc$, a finite action alphabet $\Ac$ and a collection of conditional pmfs $p(y|x,s)$ on $\Yc$.
The channel is memoryless in the sense that, without feedback, $p(y^n|x^n,s^n) = \prod_{i=1}^n p_{Y|X,S}(y_i|x_i,s_i)$, and given the action sequence, the state is memoryless in the sense that $(S_1,S_2, \ldots)$ are independent
and identically distributed (i.i.d.) with $S_i \sim p_S(s_i|a_i)$.

\begin{figure}[t]
\centering
\includegraphics[scale=0.25]{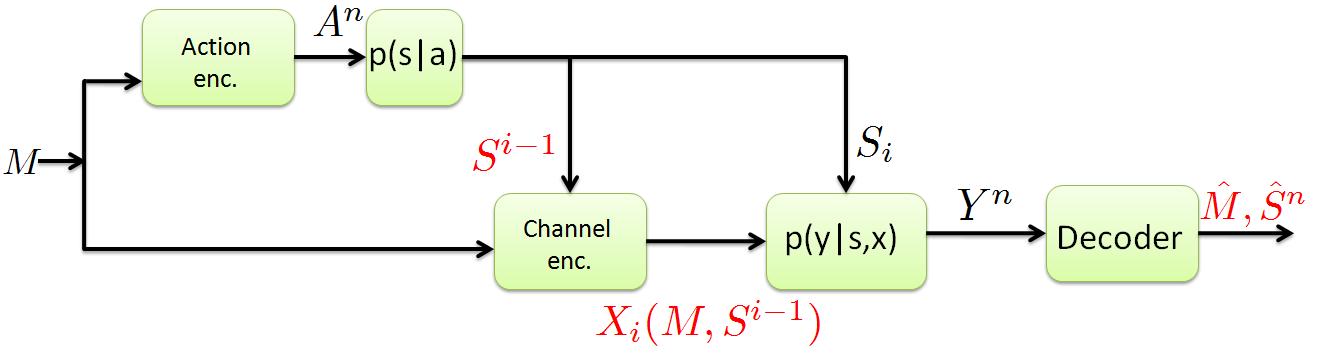}
\label{fig:action_state}
\caption{Strictly causal action dependent state communication.}
\end{figure}

A $(2^{nR}, n)$ code for strictly causal action dependent state communication consists of
\begin{itemize}
\item a \underline{message set} $[1::2^{nR}]$,
\item an \underline{action encoder} that assigns an action sequence $a^n(m)\in \Ac^n$ to each message $m \in [1::2^{nR}]$  
\item a \underline{channel encoder} that assigns a symbol $x_i(m,s^{i-1})\in \Xc$ to 
each message $m \in [1::2^{nR}]$ and past state sequence $s^{i-1} \in \Sc^{i-1}$ for $i \in [1::n]$, and
\item a \underline{decoder} that assigns a \textit{message estimate} $\mh \in [1::2^{nR}]$ (or an error message $\error$)
and a \textit{state sequence estimate} $\sh^n \in \hat\Sc^n$ to each received sequence $y^n \in \Yc^n$.
\end{itemize}

We assume that $M$ is uniformly distributed over the message set. The average probability of error 
is defined as $\pen = \P\{\Mh \ne M\}$. The fidelity of the state estimate is measured by the \emph{expected distortion}
\[
\E(d(S^n,\Sh^n)) = \frac{1}{n}\sum_{i=1}^n \E(d(S_i,\Sh_i)),
\]
where $d: \Sc \times \hat\Sc \to [0,\infty)$ is a distortion measure
between a state symbol $s \in \Sc$ and a reconstruction symbol $\sh \in \hat\Sc$. Without loss of generality, we assume that for every symbol $s \in \Sc$ there exists a reconstruction symbol $\sh \in \hat\Sc$ such that $d(s,\sh) = 0$. A rate--distortion pair is said to be achievable
if there exists a sequence of $(2^{nR},n)$ codes such that $\lim_{n\to \infty} \pen = 0$
and $\limsup_{n\to \infty} \E d(S^n,\Sh^n) \le D$.
The capacity--distortion function $\CSC^A(D)$ is defined as in~\cite{Zhang--Vedantam--Mitra2011} and is the supremum of the rates $R$ such that $(R,D)$ is achievable.

We characterize this optimal tradeoff between information transmission rate (capacity $C$)
and state estimation (distortion $D$) as follows.

\begin{theorem}\label{thm:cap_dis_action_strict_causal}
The capacity--distortion function for \emph{strictly causal} action dependent state communication is
\begin{equation*}
\CSC^A(D) = \max \bigl(I(U, A, X;Y)-I(U, X;S|A)\bigr),
\end{equation*}
where the maximum is over all conditional pmfs $p(a)p(x|a)p(u|x,s,a)$ and function $\sh(u,x,a,y)$ such that $\E(d(S,\Sh)) \le D$ and $I(U, X;Y|A)-I(U, X;S|A)\geq 0$. 
\end{theorem}

\begin{remark}\label{rem:nondegraded}
We observe that $\CSC^A(D)$ remains unchanged for seemingly for general channels of the form $p(y|s,x,a)$. This fact can be shown directly by defining a new state $S'=(S,A)$ and applying the above characterization.
\end{remark}

\begin{remark}
When both the sender and the receiver is oblivious of the channel state, the capacity--distortion function for action dependent state communication can be obtained by choosing $U=\emptyset$ and is given by,
\begin{align*}
C^A(D) & = \max I(X,A;Y),
\end{align*}
where the maximum is over all conditional pmfs $p(a)p(x)$ and function $\sh(x,a,y)$ such that $\E(d(S,\Sh)) \le D$. 
\end{remark}

Before proving the Theorem \ref{thm:cap_dis_action_strict_causal}, we recall a lemma from~\cite{Choudhuri--Kim--Mitra2010} and summarize a few useful properties of $\CSC^A(D)$ (similar to the~\cite[Corollary~1]{Choudhuri--Kim--Mitra2010},\cite{Zhang--Vedantam--Mitra2011}), which will be useful in proving the converse. 

\begin{lemma}\label{lem:data_process}
Suppose $Z \to V \to W$ form a Markov chain
and $d(z,\zh)$ is a distortion measure. Then for every reconstruction function $\zh(v,w)$, there exists
a reconstruction function $\zh^*(v)$ such that
\[
\E \bigl[d(Z,\zh^*(V))\bigr] \le \E \bigl[d(Z, \zh(V,W))\bigr].
\]
\end{lemma}

This lemma traces back to Blackwell's notion of channel 
ordering~\cite{Blackwell1953, Raginsky2011} and can be interpreted as a \textit{data processing inequality for estimation}.  

\begin{corollary}\label{cor:cap_dis_action_strict_causal}
The capacity-distortion function $\CSC^A(D)$ in Theorem \ref{thm:cap_dis_action_strict_causal}  has the following properties:\\
(1) $\CSC^A(D)$ is a non-decreasing concave function of $D$ for all $D\geq D^*$,\\
(2) $\CSC^A(D)$ is a continuous function of $D$ for all $D > D^*$,\\
(3) $\CSC^A(D^*)=0$ if $D^*\neq 0$ and $\CSC(D^*)\geq 0$ if $D^*= 0$,\\
where $D^*$ is the minimum distortion with strictly causal channel state at the sender akin to the zero rate case in~\cite{Choudhuri--Kim--Mitra2010}. 
\end{corollary}


\subsection{Sketch of Achievability:} 
We use $b$ transmission blocks, each consisting of $n$ symbols. The channel encoder uses a rate-splitting technique, whereby in block $j$, it appropriately allocates it's rate between cooperative transmission of common message $m_j$ and a description of the state sequence $S^n(j-1)$ in block $j-1$. Typical sets are defined as in~\cite{Orlitsky--Roche2001}. 

\medskip\noindent\textbf{Codebook generation.}
Fix a conditional pmf $p(a)p(x|a)$$p(u|x,s,a)$ and function $\sh(u,x,y,a)$ that attain $\CSC^A(D/(1+\e))$, where $D$ is the desired distortion, and let $p(u|x,a) = \sum_s p(s|a) p(u|x,s,a)$. For each $j \in [1::b]$,
randomly and independently generate $2^{nR}$ sequences $a^n(m_j)$, $m_j\in [1::2^{nR}]$,
each according to $\prod_{i=1}^n p_A(a_i)$ and for each $a^n(m_j)$, generate 
$2^{nR_S}$ sequences $x^n(m_j, l_{j-1})$, $m_j\in [1::2^{nR}], l_{j-1} \in [1::2^{nR_S}]$,
each according to $\prod_{i=1}^n p_X(x_i|a_i)$.
For each $m_j\in [1::2^{nR}], l_{j-1} \in [1::2^{nR_S}]$, 
randomly and conditionally independently generate
$2^{n\Rt_S}$ sequences $u^n(k_j|m_j,l_{j-1})$, $k_j \in [1::2^{n\Rt_S}]$, 
each according to $\prod_{i=1}^n p_{U|X,A}(u_i|x_i(m_j,l_{j-1}),a_i(m_j))$.
Partition the set of indices $k_j \in [1::2^{n\Rt_S}]$ 
into equal-size bins $\Bc(l_j) = [(l_j-1)2^{n(\Rt_S-R_S)}+1::l_j2^{n(\Rt_S-R_S)}]$,
$l_j \in [1::2^{nR_S}]$. The codebook is revealed to the both encoder and the decoder.

\medskip\noindent\textbf{Encoding.} 
By convention, let $l_0 = 1$. At the end of block $j$, 
the sender finds an index $k_j$ such that 
\[
(s^n(j), u^n(k_j|m_j,l_{j-1}), x^n(m_j,l_{j-1}),a^n(m_j)) \in \aepvar.
\]
If there is more than one such index, it selects one of them uniformly at random.
If there is no such index, it selects an index from $[1::2^{n\Rt_S}]$ uniformly at random. In block $j+1$, the action encoder chooses the action sequence $a^n(m_{j+1})$, where $m_{j+1}$ is the new message index to be sent in block $j+1$. Let $s^n(j+1)$ be the channel state sequence generated in response to the action sequence. The channel encoder then transmits $x^n(m_{j+1}, l_j)$ over the state dependent channel in block $j+1$, where $l_j$ is the bin index of $k_j$. 

\medskip\noindent\textbf{Decoding.}
Let $\e > \e'$. At the end of block $j+1$, the receiver finds the unique index $\mh_{j+1}, \lh_j$
such that $(x^n(\mh_{j+1}, \lh_j), y^n(j+1),a^n(\mh_{j+1})) \in \aep$. It then looks for the unique compression index $\kh_j \in \Bc(\lh_j)$ such that $(u^n(\kh_j|\mh_j, \lh_{j-1}), x^n(\mh_j, \lh_{j-1}), a^n(\mh_j), y^n(j)) \in \aep$ and $\kh_j\in \Bc(\lh_j)$. Finally it computes the reconstruction sequence as $\sh_i(j) = \sh(u_i(\kh_j|\mh_j, \lh_{j-1}),x_i(\mh_j, \lh_{j-1}),a_i(\mh_j),y_i(j))$ for $i \in [1::n]$.

Note that the achievablity scheme resembles the one in~\cite{Choudhuri--Kim--Mitra2010}, as in conditioned on the action sequence $a^n(m)$, we use a similar block Markov strategy to convey the state to the decoder. So essentially the action sequence adds one more degrees of freedom to the framework of~\cite{Choudhuri--Kim--Mitra2010}.   


\subsection{Proof of the Converse}
\par We need to show that given any sequence of $(2^{nR}, n)$-codes with $\lim_{n\to \infty} \pen = 0$ and $\E (d(S^n, \Sh^n))\le D$, we must have $R\le \CSC^A(D)$. We identify the auxiliary random variables $U_i:=(M,S^{i-1},Y_{i+1}^n,A^{n\backslash i})$, $i \in [1::n]$ with $n\backslash i=[1::n]-{i}$ and $(S_0, Y_{n+1})=(\emptyset, \emptyset, \emptyset)$. Note that, as desired, $U_i \to (X_i,S_i) \to Y_i$ form a Markov chain. Consider
\begin{align*}\label{eq:converse1}
nR & = H(M)\nonumber\\
& \stackrel{(a)}{\leq} I(M;Y^n)+n\epsilon_n\nonumber\\
& = \sum_{i=1}^n I(M;Y_i|Y_{i+1}^n)+n\epsilon_n\nonumber\\
& \leq \sum_{i=1}^n I(M,Y_{i+1}^n;Y_i)+n\epsilon_n\nonumber\\
& = \sum_{i=1}^n (I(M,Y_{i+1}^n,S^{i-1};Y_i) - I(S^{i-1};Y_i|M,Y_{i+1}^n))\\
& +n\epsilon_n\nonumber
\end{align*}
\begin{align*}
& \stackrel{(b)}{=} \sum_{i=1}^n I(M,Y_{i+1}^n,S^{i-1},A^n;Y_i)\\
& - \sum_{i=1}^n I(Y_{i+1}^n;S_i|M,S^{i-1},A^n) +n\epsilon_n\nonumber\\
& \stackrel{(c)}{=} \sum_{i=1}^n I(M,Y_{i+1}^n,S^{i-1},A^n;Y_i)\\
& - \sum_{i=1}^n I(M,S^{i-1},Y_{i+1}^n,A^{n\backslash i};S_i|A_i) +n\epsilon_n\nonumber\\
& \stackrel{(d)}{=} \sum_{i=1}^n (I(U_i,X_i,A_i;Y_i)- I(U_i,X_i;S_i|A_i)) +n\epsilon_n,
\end{align*}
where (a) can be shown by Fano's inequality (see \cite[Theorem $7.7.1$]{Cover--Thomas2006}), (b) follows from the Csis$\acute{z}$ar sum identity~\cite[Sec.~2.3]{El-Gamal--Kim2011} and since $A^n$ is a function of $M$, (c) follows from the fact that given $A_i$, $(M,S^{i-1},A^{n\backslash i})$ is independent of $S_i$, and (d) is true as $X_i$ is a function of $(M,S^{i-1})$. Similarly, for this choice of $U_i$,
\begin{align*}
\sum_{i=1}^n I(U_i,X_i;S_i|A_i) & = \sum_{i=1}^n I(M,S^{i-1},Y_{i+1}^n,A^{n\backslash i},X_i;S_i|A_i)\\
& = \sum_{i=1}^n I(Y_{i+1}^n;S_i|M,S^{i-1},A^n)\\
& \stackrel{(b)}{=} \sum_{i=1}^n I(S^{i-1};Y_i|M,Y_{i+1}^n, A^n)\\
& \leq \sum_{i=1}^n I(M, S^{i-1},Y_{i+1}^n, A^{n\backslash i};Y_i|A_i)\\
& \stackrel{(d)}{=} \sum_{i=1}^n I(U_i, X_i;Y_i|A_i).
\end{align*}
So now we have 
\begin{align*}
R & \le \frac{1}{n} \sum_{i=1}^n I(U_i,X_i,A_i;Y_i) - \sum_{i=1}^n I(U_i,X_i;S_i|A_i) +n\epsilon_n\nonumber\\
& \stackrel{(a)}{\le} \frac{1}{n} \sum_{i=1}^n \CSC^A(\E(d(S_i, \sh_i(U_i,X_i,A_i,Y_i))))+n\epsilon_n\nonumber\\
& \stackrel{(b)}{\le} \CSC^A\bigl(\frac{1}{n} \sum_{i=1}^n \E(d(S_i, \sh_i(U_i,X_i,A_i,Y_i)))\bigr)+n\epsilon_n\nonumber\\
& \stackrel{(c)}{\le} \CSC^A(D),
\end{align*}
where (a) follows from the definition of capacity-distortion function, (b) follows by the concavity of $\CSC^A(D)$ (see Property $1$ of Corollary \ref{cor:cap_dis_action_strict_causal}), and (c) can be shown using Lemma~\ref{lem:data_process} and Corollary \ref{cor:cap_dis_action_strict_causal}. This completes the proof of Theorem \ref{thm:cap_dis_action_strict_causal}. Note that main difficulty of the converse is to identify $U_i$, which not only has to satisfy the rate and distortion condition (as in~\cite{Choudhuri--Kim--Mitra2011}), but also need to satisfy the additional information inequality. 


\section{Adaptive Action}

It is natural to wonder whether ``feedback'' from the past states at the action stage ($a_i(m,s^{i-1})$) increases the capacity-distortion function or not. For an extreme example, consider a channel for which $p(y|s,x,a)=p(y|s,a)$. Clearly, the capacity--distortion function for any such channel with only message dependent non-adaptive action ($a^n(m)$) is same as that of no CSI, since the action encoder is oblivious of the channel state. But with adaptive action, the action encoder can perform a block Markov strategy to yield a potentially larger capacity--distortion function, which is summarized below without proof.

\begin{theorem}\label{thm:cap_dis_adap_action_strict_causal}
The capacity--distortion function for \emph{strictly causal adaptive} action dependent state communication is
\begin{equation*}
\CSC^{AA}(D) = \max \bigl(I(U, A, X;Y)-I(U, X, A;S)\bigr),
\end{equation*}
where the maximum is over all conditional pmfs $p(a)p(x|a)p(u|x,s,a)$ and function $\sh(u,x,a,y)$ such that $\E(d(S,\Sh)) \le D$. 
\end{theorem} 

Note that the \textit{unconstrained capacity} remains \textit{unchanged} even if we allow the actions to depend on the past states. In general, $\CSC^{AA}(D)\geq \CSC^{A}(D)$ as the adaptive action helps the receiver to get a better estimate of the state. Finally, by setting $A = \emptyset$ in Theorem~\ref{thm:cap_dis_adap_action_strict_causal}, we recover the result by~\cite{Choudhuri--Kim--Mitra2010} on the capacity--distortion function when the i.i.d. state information is available strictly causally at the encoder. 

\begin{remark}
When the past states are available at both the encoders, the encoders cooperate to send information consisting of the common message and a description of the state in previous block (similar to sending a common message over multiple access channel (MAC)), whereas in the non-adaptive action scenario, while the common message is sent cooperatively, description of the \textit{state} is a \textit{private} message of the channel encoder.
\end{remark}
\section{Illustrative Examples}
In the following subsections, we illustrate Theorem~\ref{thm:cap_dis_action_strict_causal} and Theorem~\ref{thm:cap_dis_adap_action_strict_causal} through examples.

\subsection{Actions Seen by Decoder:}
Consider the case where the decoder also has access to the actions taken. Noting that this is a special case of our
setting by taking the pair $(Y, A)$ as the new channel output, that $U\to (X,S,A) \to Y$ if and only if $U\to (X,S,A) \to (Y,A)$. We obtain that the capacity--distortion function for the case of message depepdent action is given by
\[
\CSC^{A}(D) = \max \bigl(H(A) + I(U, X;Y|A)-I(U, X;S|A)\bigr),
\]
where the maximization is over the same set of distributions and same feasible set as in Theorem~\ref{thm:cap_dis_action_strict_causal}. Similarly we can evaluate the capacity--distortion function for the case of adaptive actions. This expression is quite intuitive: The amount
of information per symbol that can be conveyed through the actions in the first stage is represented by the term $H(A)$. In
the second stage, both encoder and decoder know the action sequence, so they can condition on it and can perform the usual block Markov strategy on each subsequence associated with each action symbol, achieving a rate of $I(U, X;Y|A)-I(U, X;S|A)$. The maximization is a search for the optimal tradeoff between the amount of information that can be conveyed by the actions,
and the quality of the second stage channel that they induce.

\subsection{Gaussian Channel with Additive Action Dependent State}
Consider the Gaussian channel with additive action dependent state~\cite{Weissman2010}
\begin{align*}
Y & = X + S + Z = X + A +\tilde{S} +Z,
\end{align*}
where $\tilde{S} \sim \N(0,Q)$ and the noise $Z \sim \N(0,N)$ are independent.
Assume an expected average power constraint on both the channel and action encoder 
\[
\sum_{i=1}^n \E(x_i^2(m,S^{i-1})) \le nP_X,\sum_{i=1}^n \E(a_i^2) \le nP_A.
\]
We consider the \emph{squared error (quadratic) distortion measure} $d(s,\sh) = (s-\sh)^2$. When the action sequnce is only a function of the message, using Theorem~\ref{thm:cap_dis_action_strict_causal} we have the following.

\begin{proposition}\label{prop:Gaussian_action_state_trade_off}
The capacity--distortion function of the Gaussian channel with message dependent action is
\[
\CSC^A(D) =
\begin{cases}
0, & 0 \le D < D_{min}^A,\\
\frac{1}{2}\log\left(\frac{P^A}{QN/D}\right), & D_{min}^A \le D < D_{max},\\
\C\left(\frac{(\sqrt{P_X}+\sqrt{P_A})^2}{Q+N}\right), & D \ge D_{max}.
\end{cases}
\]
where $\C(x)=\log(1+x)$, $D_{min}^A=\frac{QN}{P_X+Q+N}$, $D_{max}=\frac{QN}{Q+N}$ and $P^A=P_X+Q+N+P_A+2\sqrt{P_A(P_X-(\frac{QN}{D}-(Q+N)))}$.
\end{proposition}

When we allow the action encoder to observe the past states (adaptive action), the capacity--distortion follows from Theorem~\ref{thm:cap_dis_adap_action_strict_causal} and it has the similar form of Proposition~\ref{prop:Gaussian_action_state_trade_off}, but $P^A$ and $D_{min}^{A}$ are replaced by $P^{AA}$ and $D_{min}^{AA}$, respectively, where $P^{AA}=P_X+Q+N+P_A+2\sqrt{P_AP_X}$ and $D_{min}^{AA}=QN/P^{AA}$. 

\begin{figure}[t]
\centering
\includegraphics[scale=0.35]{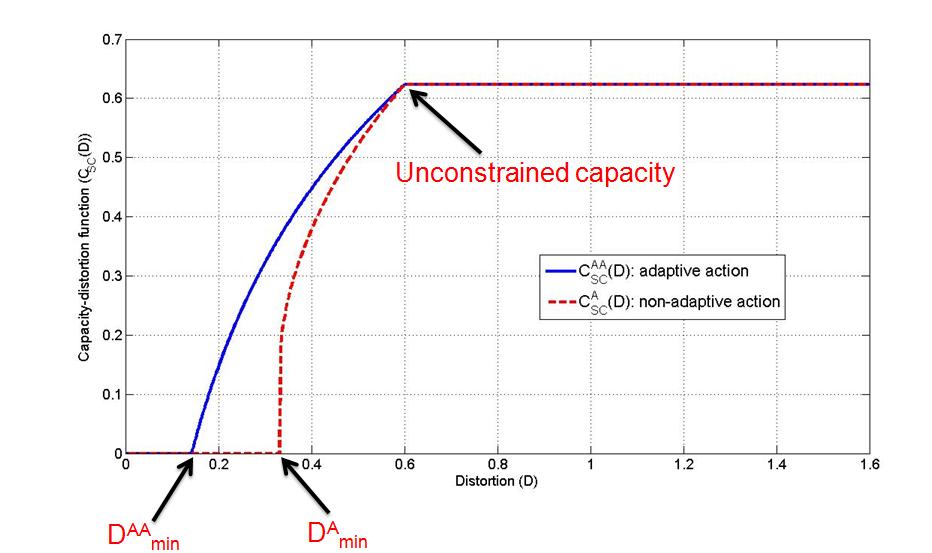}
\label{fig:adapvsnonadap}
\caption{Capacity--distortion function: adaptive vs. non-adaptive}
\end{figure}

The proof of the proposition is omitted here for brevity. Note that since $P^{AA}\geq P^A$, the capacity--distortion function is larger in the adaptive action scenario (see Figure~\ref{fig:adapvsnonadap}). In fact, the minimum distortion achievable with adaptive action is smaller than that of non-adaptive action. But the unconstrained capacity (capacity--distion function for $D\ge D_{max}$) is same in both the cases, which implies that adaptive action in useful in estimation rather than in information transmission. Finally by substituting $P_A=0$, both the capacity--distortion functions reduces to that in~\cite{Choudhuri--Kim--Mitra2010}. 

\begin{figure}[t]
\centering
\includegraphics[scale=0.30]{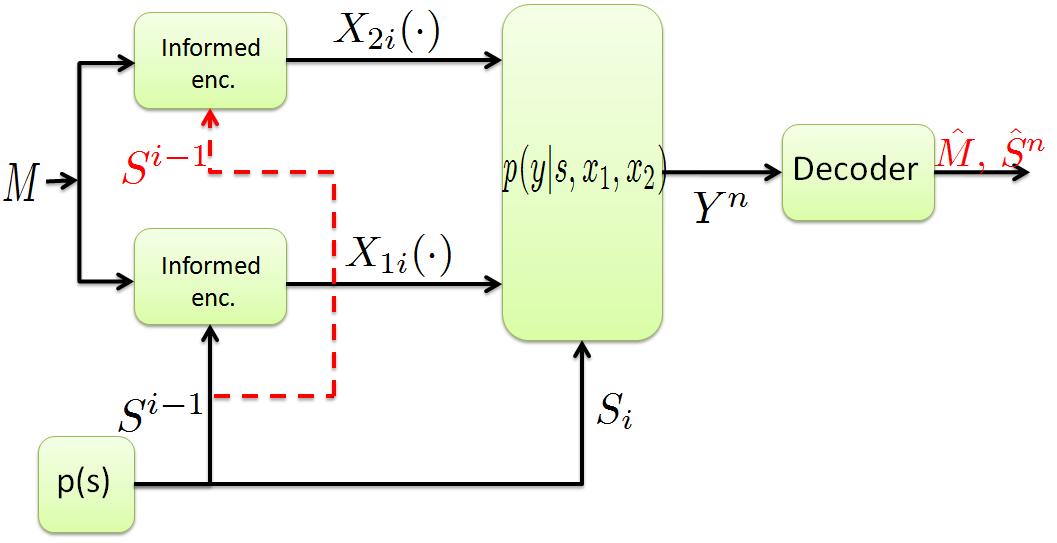}
\label{fig:action_state3}
\caption{State dependent MAC with strictly causal CSI at both encoders.}
\end{figure}

\subsection{State dependent MAC}
Consider communicating a common message over a memoryless state-dependent MAC (see Figure \ref{fig:action_state3}) characterized by $p(y|s,x_1,x_2)$, where the state sequence is known strictly-causally to both encoders. This problem can be seen as a special case of our adaptive action setting via the following associations:
\[
A=X_2, X=X_1, p(s|a)=p(s), p(y|s,a,x)=p(s,x_1,x_2).
\]
Applying Theorem~\ref{thm:cap_dis_adap_action_strict_causal} to this case, keeping in mind the Remark~\ref{rem:nondegraded} following the statement of the Theorem~\ref{thm:cap_dis_action_strict_causal}, regarding channels of the form $p(y|s,x,a)$, we get that the capacity--distortion function is given by
\[
\CSC^{S}(D) = \max \bigl(I(U, X_2, X_1;Y)-I(U, X_1;S|X_2)\bigr),
\]
where the maximum is over $p(x_1,x_2)p(u|x_1,s,x_2)$ and function $\sh(u,x_1,x_2,y)$ such that $\E(d(S,\Sh)) \le D$. This setting was considered in \cite{Lapidoth--Steinberg2010, Li--Simeone--Yener2011} and it recovers the common message capacity results of \cite{Lapidoth--Steinberg2010, Li--Simeone--Yener2011}. One can also consider a scenario where the state sequence is known strictly-causally to the first encoder, but unknown at the second encoder and at the receiver. This problem, motivated by multiterminal communication scenarios involving transmitters with different degrees of channel state information, is a special case of Theorem~\ref{thm:cap_dis_action_strict_causal}. We can show that the capacity--distortion function ($\CSC^{AS}(D)$) is the same as $\CSC^{S}(D)$ with the additonal constraint of $I(U, X_1;Y|X_2)-I(U, X_1;S|X_2)\geq 0$ on the feasible distributions. Clearly $\CSC^{S}(D)\geq \CSC^{AS}(D)$, since with symmetric channel state information, the encoders can jointly perform both message and state cooperation as opposed to only message cooperation when the state information is available at only one of the encoders.


\section{Conclusions}
Motivated by an active classification problem with autonomous vehicles, we combine the frameworks of \cite{Weissman2010} and \cite{Choudhuri--Kim--Mitra2010}, to examine the problem wherein the formation of channel states is affected by actions taken at the encoder; further, the decoder has the two simultaneous goals of estimating the channel state up to some distortion and simultaneously decoding the transmitted message.  We characterize the capacity-distortion function for this problem where the channel states are known strictly causally at (a) only the channel encoder, and (b) both the action encoder and channel encoder.  By realizing that, conditioned on the action sequence, our framework is similar to that in \cite{Choudhuri--Kim--Mitra2010}, we have shown that a two stage encoding strategy is optimal.  In the first stage, the action is communicated and in the second stage, conditioned on the action sequence, a block Markov strategy is performed to utilize the strictly causal CSI at the encoder(s). We have also shown that the state-dependent MAC with symmetric and asymmetric state information is a special case of our framework and thus are able to use our results to recover the common message capacity results of the MAC with strictly causal CSI (see~\cite{Lapidoth--Steinberg2010, Li--Simeone--Yener2011}).       


\bibliographystyle{IEEEtran}
\bibliography{nit}
             
\end{document}